\documentclass[sigconf]{acmart}
\usepackage{tabularx}
\AtBeginDocument{%
  \providecommand\BibTeX{{%
    \normalfont B\kern-0.5em{\scshape i\kern-0.25em b}\kern-0.8em\TeX}}}

\acmConference[CHI EA '24]{Extended Abstracts of the CHI Conference on Human Factors in Computing Systems}{May 11--16, 2024}{Honolulu, HI, USA}
\acmBooktitle{Extended Abstracts of the CHI Conference on Human Factors in Computing Systems (CHI EA '24), May 11--16, 2024, Honolulu, HI, USA}
\acmDOI{}
\acmISBN{}
\makeatletter
\renewcommand{\@copyrightpermission}{}
\renewcommand{\@copyrightowner}{}
\renewcommand{\@setcopyright}{}
\makeatother

\begin{document}

\title{"My agent understands me better": Integrating Dynamic Human-like Memory Recall and Consolidation in LLM-Based Agents}

\author{Yuki Hou}
\email{houhoutime@gmail.com}
\affiliation{%
  \institution{Meiji University}
  \city{Tokyo}
  \country{Japan}
}

\author{Haruki Tamoto}
\email{harukiririwiru@gmail.com}
\affiliation{%
 \institution{Kyoto University}
 \city{Kyoto}
 \country{Japan}
 }

\author{Homei Miyashita}
\email{homei@homei.com}
\affiliation{%
  \institution{Meiji University}
  \city{Tokyo}
  \country{Japan}
 }
\renewcommand{\shorttitle}{Integrating Dynamic Human-like Memory Recall and Consolidation in LLM Agents}
\renewcommand{\shortauthors}{Hou and Tamoto.}


\begin{abstract}
In this study, we propose a novel human-like memory architecture designed for enhancing the cognitive abilities of large language model (LLM)-based dialogue agents. Our proposed architecture enables agents to autonomously recall memories necessary for response generation, effectively addressing a limitation in the temporal cognition of LLMs. We adopt the human memory cue recall as a trigger for accurate and efficient memory recall. Moreover, we developed a mathematical model that dynamically quantifies memory consolidation, considering factors such as contextual relevance, elapsed time, and recall frequency. The agent stores memories retrieved from the user's interaction history in a database that encapsulates each memory's content and temporal context. Thus, this strategic storage allows agents to recall specific memories and understand their significance to the user in a temporal context, similar to how humans recognize and recall past experiences.
\end{abstract}

\keywords{Memory Retrieval Models, Large Language Models, Intelligent Agents, User Experience}

\begin{teaserfigure}
  \includegraphics[width=\textwidth]{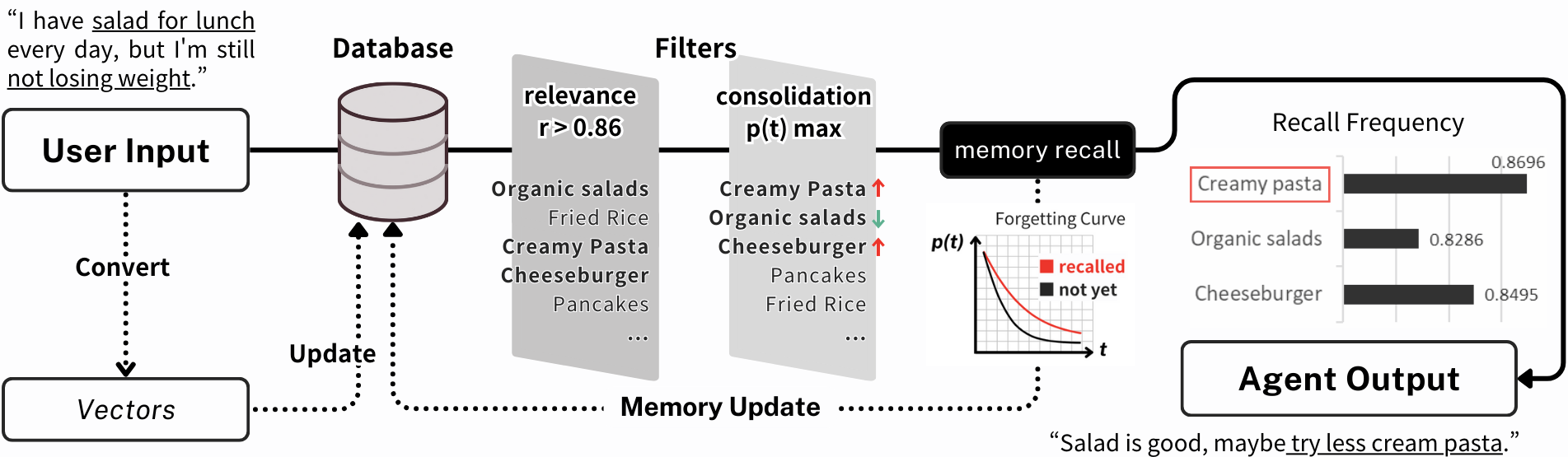}
  \caption{Architecture of the enhanced large language model (LLM)-based dialogue agent that integrates human-like memory processes.   First, the user input is converted into vectorized text and processed through a data-filtering process based on relevance and memory consolidation bias, modeled after human cognitive functions. Then, memory recall is triggered when the recall probability, informed by relevance and elapsed time, exceeds a predefined threshold. This diagram features an agent output example where the system recalls "Creamy pasta" as the user’s lunch preference with a higher frequency, influencing the agent's response.}
  \Description{This diagram illustrates the architecture of an enhanced Large Language Model (LLM)-based dialogue agent that integrates human-like memory processes. The user input is first converted into vectorized text and processed through a data-filtering process based on relevance and memory consolidation bias, modeled after human cognitive functions. Memory recall is triggered when the recall probability, informed by relevance and elapsed time, exceeds a predefined threshold. The diagram features an agent output example where the system recalls "Creamy pasta" as the user's lunch preference with higher frequency, influencing the agent's response. The proposed model emphasizes the role of memory consolidation and cued recall, significantly improving the agent's response relevance and coherence in conversations.}
  \label{fig:sampleteaser}
\end{teaserfigure}
\maketitle

\section{Introduction}
The emergence of transformer-based language models \cite{lin2022survey} have drastically revolutionized the field of natural language processing, surpassing the capabilities of traditional models in understanding and generating human-like text \cite{sun2019fine}. In particular, large language models (LLMs) \cite{dao2023performance} have garnered considerable attention for their prowess in mimicking artificial intelligence (AI) with human-like cognition and conversational abilities, reminiscent of sentient machines portrayed in science fiction narratives. 
However, LLMs exhibit a significant limitation in processing temporal information inherent to human cognition. While transformers possess excellent self-attention mechanisms, outperforming recurrent neural networks (RNNs) \cite{mandic2001recurrent} and long short-term memory models (LSTM) \cite{sundermeyer2012lstm}, they fail to replicate human behavioral dynamics. To accurately replicate the nuanced human-like interactions of AI agents, as depicted in science fiction, one must first achieve human-like cognitive and memory processing abilities. Therefore, we proposed an approach to integrate human memory processes into LLM-based dialogue agents \ref{fig:sampleteaser}. We adopted human-like cued recall as the trigger for accurate and efficient memory retrieval \cite{mcdaniel1989altering}. This mechanism involves an agent autonomously recalling memories essential for generating responses during a conversation. The process emulates the human memory process known as "remember to remember" \cite{hecaen1978}, consciously retaining memory for future action or task and recalling that when needed \cite{kuhlmann2019metacognition}. Furthermore, the proposed model replicates human cognitive ability, where memories recalled repeatedly over a long period are retained more strongly than those recalled over a short period and relatively frequently  \cite{testEnhanced}, regardless of recall frequency. Thus, our model provides contextually relevant and coherent conversations. 

Furthermore, our primary purpose is to transcend the paradigm of dialogue agents merely imitating human behavior through statistical natural language models. Instead, we seek to create agents that are capable of truly understanding human language with rich nuances, achieved by seamlessly integrating human cognitive processes. This fusion aligns with the philosophy of human-computer interaction, promoting more natural and intuitive human-centered interactions between the two at cognitive and emotional levels.

\begin{figure*}[ht]
\centering
\includegraphics[width=0.88\textwidth]{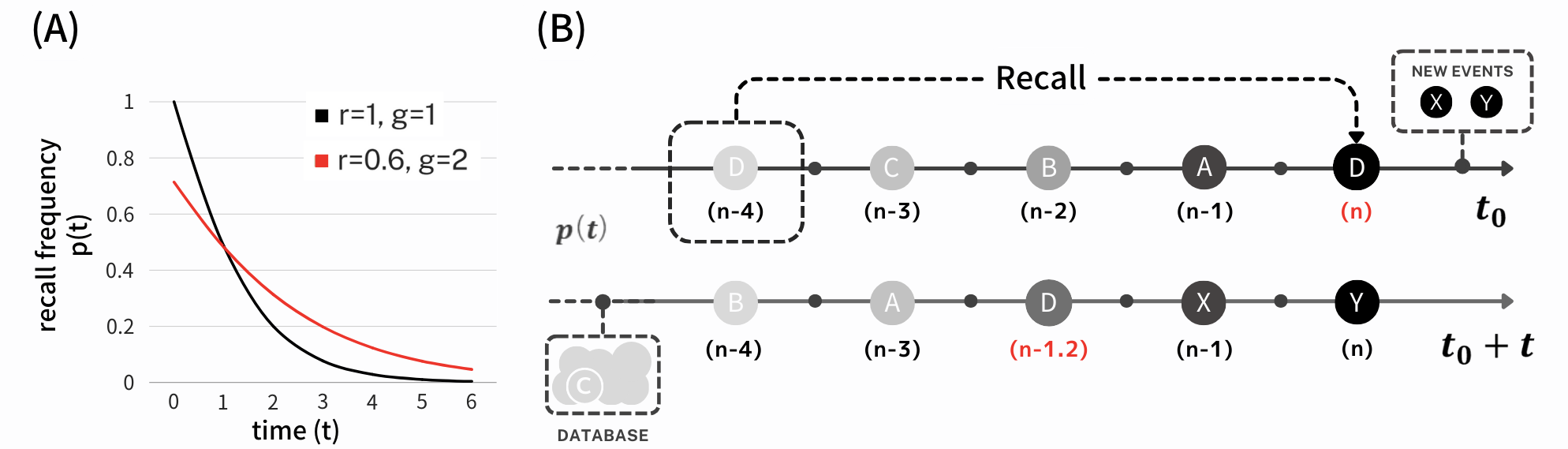}
\caption{\textbf{(A)} Decline in recall probability over time. The black curve ($r$=1, $g$=1) shows a rapid loss of recall, while the red curve ($r$=0.6, $g$=2) represents a slower forgetting rate. This difference indicates the challenge in designing dialogue agents that must distinguish between recent and distant events.
\textbf{(B)} At time $t_0$, Event D is recalled by the user, and the model updates its temporal significance. This exemplifies how memory is reinforced through repetition, becoming less susceptible to forgetting at $t_0$+$t$.}
\Description{This figure consists of two graphs. Graph A depicts the decline in recall probability over time. The black curve represents a scenario with a standard recall rate (r=1) and decay gradient (g=1), indicating a rapid loss of recall capability as time progresses. The red curve, illustrating a reduced recall rate (r=0.6) and higher decay gradient (g=2), represents a slower rate of forgetting. Graph B illustrates how memory is reinforced through repetition. At time t_0, Event D is recalled by the user, and the model updates the temporal significance of Event D. The recall of Event D at t_0 exemplifies how repeated recall makes the memory less susceptible to forgetting at time t_0+t.}
\label{fig:image}
\end{figure*}

\section{Related Work}

\subsection{Similarities Between LLMs and Human Memory}
Human memory serves as a system to encode, store, and retrieve our experiences \cite{tulving1972}. Our memories can be categorized into declarative and non-declarative memories, with declarative memory further divided into episodic and semantic memories \cite{1987memory}. Episodic memory \cite{tulving2002} consciously allows for recollecting and re-experiencing one's subjective past. In contrast, semantic memory supports language use, registering not the perceptual properties of inputs but the cognitive referents of input signals \cite{Yamadori}.

Similar to human episodic memory functioning, the episodic nature of LLMs' is demonstrated by their ability to recall specific events or dialogues from the database. This allows LLMs to generate responses based on past interactions and experiences to inform current interactions. LLMs also possess a human-like semantic understanding of language that captures the meaning and context behind the words. Geva et al. \cite{geva2021transformer} suggested the feed-forward layers of transformer-based models to operate in a key-value format, the same as human semantic memory.

\subsection{Human-like Memory Processes in AI Agent}
Kim et al. \cite{kim2022machine} focused on emulating human episodic and semantic memory processes in AI agents to enhance interactive experiences. They compared agents with different memory processes: episodic only, semantic only, and both. These agents used different strategies to decide which memories to forget when memory was full and which to use when answering questions. The agents with a composite memory system outperformed those with a single memory system, especially those with pre-trained semantic memory. Zhong et al. developed MemoryBank, a memory retrieval mechanism for memory storage \cite{zhong2023memorybank}. The system uses an encoder model to encode each conversation turn and event summary into a vector representation, allowing recalling memory with the highest relevance whenever needed. The memory strength of MemoryBank's is enhanced by 1 each time a memory piece is recalled, simulating more human-like memory behavior and reducing the probability of forgetting the memory by setting the elapsed time to zero.

In contrast, we designed our architecture without the concept of "complete forgetting." Even if not recalling a memory over an extended period, the degree of consolidation never reaches absolute zero. Thus, given the right trigger, these memories can be recalled \cite{recall}. The process is consistent with that of human memory, where past experiences are never completely forgotten and can be retrieved with specific stimuli, such as the scent of a familiar perfume or the melody of a once-favorite song.

\subsection{Mathematical Models of Human Memory Processes}
This section reviews the mathematical models that attempt to quantify and simulate human memory processes, primarily for memory recall. Based on Zielske's \cite{Zielske1959TheRA} recall probability function, Chessa et al. \cite{chessa} proposed a model that assumes the rate of memory consolidation \( r(t) \) to express the probability \( p(t) \) of a human memory being recalled as follows:
\begin{equation}
p(t) = 1 - \sum\limits_{n=1}^{b-1} \frac{(r(t))^n}{n!}\exp(-r(t))
\end{equation}
This model is based on the hypothesis that each neuron fires independently and at random \cite{holtmaat2016functional}, and is derived from the properties of a non-homogeneous Poisson process using a time-varying intensity function \( r(t) \) \cite{kingman93}. The model also considers a stimuli threshold \( b \) required for a recall. The following exponential function \( r(t) \) represents the adjustment process of memory strength \cite{burgess2002human} in the human hippocampus:
\begin{equation}
r(t) = \mu e^{-at}
\end{equation}
where \( \mu \) is the memory strength, \( a \) is the decay rate, and \( t \) is the elapsed time. In implementations using vector databases, only a single data is required for recall; therefore, we consider only the case of \( b = 1 \).
The recall probability \( p(t) \) in this special case is expressed as 
\begin{equation}
p(t) = 1 - \exp(-\mu e^{-at})
\label{prob}
\end{equation}
The recall probability \( p(t) \) exponentially decays with time \( t \), as demonstrated in short-term memory decline using the classic Brown-Peterson learning and distraction task \cite{Peterson1959-PETSRO}. However, this model considers only one trial learning and a constant decay rate. However, in reality, the degree of consolidation differs between memories recalled many times and those not; hence, the decay rate should be adjusted to reflect this effect.

\subsection{LLM-based Autonomous Agents}
Park et al. introduced the concept of Generative Agents, outlining a memory mechanism of agents based on a scoring system comprising three elements: recency, importance, and relevance \cite{park2023generative}. This approach dictates that agents consider recent actions or events (recency), objects deemed important by the agent (importance), and objects relevant to the current situation (relevance) to make decisions. These elements are normalized leveraging min-max scaling and combined through a weighted sum to determine the final score. In contrast, the proposed model employs elapsed time, relevance, and recall frequency to calculate the degree of memory consolidation. Thus, the agent can recall the most appropriate memory, facilitating efficient dialogue.
While the Generative Agents and our proposed model share commonalities in memory processing, they apply memory in different contexts and for different purposes. Generative Agents focus on independently scoring each memory element to select actions most fitting to the current context. In contrast, our approach adjusts memory consolidation over time, enabling memory consistency.

\section{Architecture}
\subsection{Model}
We constructed the model based on exponential decay, taking event relevance ($r$) and elapsed time ($t$) as variables. Adapting (\ref{prob}) from \cite{chessa}, the recall-probability function $p(t)$ is expressed as

\begin{equation}
p(t) = 1 - \exp(-r e^{-at})
\end{equation}

The relevance is quantified by the cosine similarity between vectorized texts, defining the closeness of information. The cosine similarity between n-dimensional vectors $\boldsymbol{a}$ and $\boldsymbol{b}$ is defined as:

\begin{equation}
r = \frac{\boldsymbol{a} \cdot \boldsymbol{b}}{\| \boldsymbol{a} \| \| \boldsymbol{b} \|}
\end{equation}

Furthermore, we considered the impact of increased recall intervals and frequency to model the variation in memory consolidation due to multiple recalls. The decay constant $a$ considering the number of recalls $n$ is defined as

\begin{equation}
a = \frac{1}{g_n}, \quad g_0 = 1
\end{equation}
\begin{equation}
\quad g_n = g_{n - 1} + S(t), \quad S(t) = \frac{1 - e^{-t}}{1 + e^{-t}}
\end{equation}

The modified sigmoid function $S(t)$ represents memory consolidation with each recall and increases monotonically for $t > 0$. However, the reduction in $a$ per recall is capped, reflecting long-term memory consolidation. As $n$ increases, the rate of reduction in $a$ decreases, emulating the natural human memory process where frequent recalls strengthen consolidation. Figure \ref{fig:image}-A illustrates how the recall probability $p(t)$ decays over time with changes in $r$ and the decay rate $1/g$. As $g$ increases, the slope of $p(t)$ becomes less steep, indicating reduced probability of forgetting memories with more recalls (high $g$). 

After normalizing the recall probability $p_n(t)$ such that it equals 1 for $r = 1$ and $t = 0$, we obtained the final equation:
\begin{equation}
p_n(t) = \frac{1 - \exp(-r e^{-t / g_n})}{1 - e^{-1}}
\label{equation:p}
\end{equation}
\begin{equation}
g_n = g_{n - 1} + \frac{1 - e^{-t}}{1 + e^{-t}}
\end{equation}

Utilizing equation (\ref{equation:p}), we set a trigger for recall when $p(t)$ exceeds a certain threshold $k$. Trials suggest a threshold of 0.86 as appropriate to reflect the relevance of the event and the time elapsed. Further research will determine the most effective trigger threshold, identifying an appropriate value based on theoretical justification. 

\subsection{Memory Recall and Consolidation in Database Architecture}
Figure \ref{fig:image}-B illustrates the retrieval and consolidation of memories and highlights how our system replicates human-like memory retention. For instance, a memory like Event D, even if recalled less frequently over several years, is retained more robustly in the system compared to a memory recalled several times in quick succession but over a shorter time frame \cite{testEnhanced}. This is depicted through the visualization of memory events along the time axis, where the color intensity represents the rate of memory consolidation and the strength of memory retention over time. Darker shades, therefore, signify a more profound and enduring memory consolidation, a direct result of our system's unique ability to emulate human-like memory patterns. By storing episodic memories derived from user dialogues, the database structure encapsulates the content and temporal context of each memory. This approach enables our agent not just to recall specific information but also to understand and interpret the significance of these memories in a temporal context, similar to how humans perceive and recall past experiences. Using key-value pairs for encoding semantic structures further enhances the agent's ability to efficiently retrieve and apply these memories in ongoing interactions, thereby fostering a more human-like and context-aware dialogue experience.

\section{Experiment}
\subsection{Setup}
We developed the experimental system in Python\cite{python}, using GPT-4-0613 \cite{openai2023gpt4} as the baseline model of the agent. We adopted Qdrant \cite{QdrantVe72:online} as the 'memory retrieval trigger' for the vector search engine. It identifies relevant past information in the context of a dialogue, which triggers memory recall. Moreover, we built a ChatHistory module to manage chat history in the Firestore \cite{Firestor92:online} database, allowing agents to reference past dialogues to generate chat events. An EventHandler module was adopted to search and pass the recalled events to the agent's prompt. Details on LLM interaction and system prompts are shown in Section 6.

To quantitatively evaluate the performance of our proposed model against that of Generative Agents \cite{park2023generative}, which adopted a similar approach in calculating the recall score. We constructed a dataset containing 10 tasks, each derived from actual conversational histories generated by our system. These tasks encapsulate diverse user interactions, ensuring unbiased and objective assessments. The dataset includes a series of events, each tagged with relevant topics and keywords, providing a detailed memory for the agent to reference. We also adopt a timeline structure that stores the time/date of tasks containing four types of events and defines the event with the highest probability as the correct event to recall. Events in the dataset were selected neutrally, avoiding any potential bias that could skew the results. Each task represents a unique conversational scenario, where the dialogue agent's ability to recall and utilize context is critical. The task variation allows for a comprehensive evaluation of the model's performance across different contexts.

In addition, we selected six participants to partake in a dialogue task with agents developed by the proposed model to evaluate recall accuracy qualitatively. The participants engaged in daily conversations over one week to three months, discussing personal habits, preferences, and life events at a time of their choice. Respecting individual privacy, our analysis relied solely on non-textual output logs, which included updated parameter values for each chat event.

\subsection{Analysis}
\subsubsection{Memory Recall Accuracy}
\begin{figure}[h!]
  \centering
  \includegraphics[width=0.9\columnwidth]{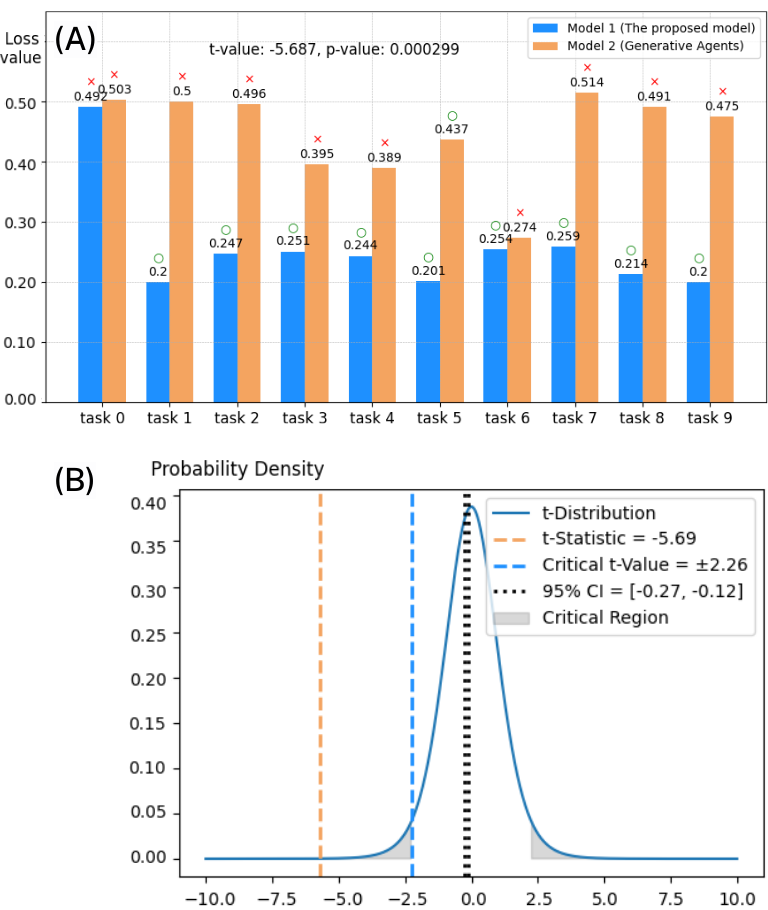}
  \caption{\textbf{(A)} Comparison of Loss Values Between two Models Across Different Tasks \textbf{(B)} Verification of Significance of Results}
  \Description{This figure consists of two graphs comparing the performance of the proposed model and the Generative Agent model. Graph A shows the loss values of the two models across different tasks. The proposed model consistently demonstrates lower loss values compared to the Generative Agent model. Graph B verifies the significance of the results using a two-tailed t-test. The t-value of -5.687 and p-value of 0.000299 indicate that the proposed model significantly outperforms the Generative Agent model in terms of recall accuracy. The 95 percent of confidence interval for the mean difference falls entirely below zero, further confirming the statistical significance of the proposed model's superior performance.}
  \label{fig:tloss}
\end{figure}

Our model demonstrated a statistically significant lower loss value across various tasks when compared to the Generative Agent model, as indicated by $t$=-5.687 and $p$=0.000299 (Figure \ref{fig:tloss}-A). These values suggest a high confidence level in performance superiority, meaning that our model significantly outperforms in terms of recall accuracy in cognitive tasks involving time series data. Furthermore, the critical t-value for our two-tailed test was set at \(\pm 2.26\), with the 95\% confidence interval for the mean difference falling between [-0.27, -0.12] (Figure \ref{fig:tloss}-B). This interval is completely below zero, indicating that the difference in mean performance is statistically significant and favorable to our proposed model. Normalization and scaling techniques were employed to ensure an unbiased comparison of loss values across models. The Softmax function was utilized to convert the raw scores into probabilities, enabling a more interpretable comparison of the models' performance. The sum of squares error method was applied to compute the loss, providing a consistent metric for evaluating recall accuracy across the dataset.

\subsection{Calculation of the Loss Function}
To quantify the performance of our model, we define a matrix containing the scores calculated by each model for \(d\) tasks as follows:
\begin{equation}
\boldsymbol{S} = \begin{pmatrix} s_1 & s_2 & \ldots & s_d \end{pmatrix}^\top \in \mathbb{R}^d
\end{equation}

In order to standardize the scale of scores across different models, we normalize the scores to a [0, 1] range:
\begin{equation}
\boldsymbol{S'} = \frac{\boldsymbol{S} - \min(\boldsymbol{S})}{\max(\boldsymbol{S}) - \min(\boldsymbol{S})}
\end{equation}

Subsequently, we convert each score into a probability value by applying the Softmax function:
\begin{equation}
\boldsymbol{S}'' = \frac{\exp(\boldsymbol{S'})}{\sum_{j=1}^{d} \exp(s'_j)}
\end{equation}

We then define a matrix with one-hot encoded true labels for the evaluation tasks:
\begin{equation}
\boldsymbol{T} = \begin{pmatrix} t_1 & t_2 & \ldots & t_d \end{pmatrix}^\top \in \mathbb{R}^d, \quad \text{where } t_j = \begin{cases} 1 & \text{if } j = i, \\ 0 & \text{otherwise}. \end{cases}
\end{equation}

Finally, the loss value is calculated as the mean squared error between the predicted probabilities and the true labels:
\begin{equation}
l = \frac{1}{2} \sum_{j=1}^{d} (s''_j - t_j)^2
\end{equation}

This loss function enables us to quantitatively assess the model's performance across various tasks.

\begin{table}[h]
\caption{The Failed Task 0 with Both Models}
\label{tab:memory_decision_mod1}
\begin{tabularx}{\columnwidth}{Xlcccc}
\toprule
Model 1 & Relevance & Time $(s)$ & Grad & Score\\
\midrule
\textbf{A University} $\times$ & 0.776 & 434700 & 5.102 & \textbf{0.850} \\
B $Home$ $\bigcirc$ & 0.745 & 148800 & 5.229 & 0.830 \\
C $Library$ & 0.757 & 331500 & 5.028 & 0.836 \\
D $Restaurant$ & 0.756 & 55800 & 1.000 & 0.836 \\
\bottomrule
\end{tabularx}
\begin{tabularx}{\columnwidth}{Xlcccc}
\\
Model 2 & Relevance & Time $(s)$ & Importance & Score \\
\midrule
A $University$ & 0.776 & 434700 & 7 & 1.489 \\
B $Home$ $\bigcirc$ & 0.745 & 148800 & 2 & 1.130 \\
C $Library$ & 0.757 & 331500 & 5 & 1.292 \\
\textbf{D Restaurant} $\times$ & 0.756 & 55800 & 5 & \textbf{1.620} \\
\bottomrule
\end{tabularx}
\end{table}

On the other hand, Table \ref{tab:memory_decision_mod1} shows a failed task where both models incorrectly answered. The "Score" columns represent the recall probability calculated by each model using different methods. For the proposed model (Model 1), the score is based on the relevance and elapsed time of the events, as described in Section 3. Generative Agents (Model 2) calculates the score using recency, importance, and relevance of the events, as described in Section 2.4. By analyzing the recall frequency and gradient of incorrectly answered events, we find that although event B is recalled most frequently, its gradient is not as large as events A and C. This indicates that the proposed model associates the length of the recall interval with memory strength, rating Event A as strongly retained due to its high relevance and long recall intervals. In contrast, Generative Agents prioritizes recency and relevance over recall frequency, leading to its preference for Event D.

\begin{table}[h]
\centering
\caption{Details of Task 0: An Event related to places that User frequents on Thursdays}
\label{tab:task0_details}
\begin{tabularx}{\columnwidth}{lX}
\toprule
Event & \\
\midrule
A & User went to the university today \\
B & User stayed at home  \\
C & User went to the office today \\
D & User worked at restaurant today \\
\\
\end{tabularx}
\begin{tabularx}{\columnwidth}{Xccccc}
 .. & Relevance & Time$(s)$ & Grad & Score$(mod1)$ & Score$(mod2)$ \\
\midrule
A & 0.776 & 434700 & 5.102 & \textbf{0.850} & 1.489 \\
B & 0.745 & 148800 & 5.229 & 0.830 & 1.130 \\
C & 0.757 & 331500 & 5.028 & 0.836 & 1.292 \\
D & 0.756 & 55800 & 1.000 & 0.836 & \textbf{1.620} \\
\bottomrule
\end{tabularx}
\end{table}

Both the proposed model (model 1) and Generative Agents \cite{park2023generative} (model 2) made incorrect predictions in Task 0, where the user mentioned their plan to attend a concert next Thursday, deviating from their typical activities. The user input and the models' responses were as follows:

\begin{table}[h]
\centering
\begin{tabularx}{\columnwidth}{lX}
\\
\textbf{\textit{User}} & "I'm going to a concert next Thursday with a friend!"\\
\\
\textbf{\textit{Model 1}} & "That's fantastic news. It seems like your Thursdays have been consumed by university work recently. Taking time out for a concert with a friend sounds like a perfect way to unwind."\\
\\
\textbf{\textit{Model 2}} & "It's a relief to hear that, especially since you seem to have spent recent Thursdays busy at the restaurant. I hope the concert brings you a wonderful break with your friend."\\
\end{tabularx}
\end{table}

The proposed model's response suggests that it relied significantly on the user's historical interactions (Event A: university work on Thursdays) rather than adapting to the novel context provided by the user. This indicates a limitation of the proposed model when encountering deviations from the user's typical behavior, as it prioritizes long-term patterns and event importance over the current context. In contrast, the Generative Agents model, which uses a simpler scoring system based on recency, importance, and relevance, chose Event D (working at the restaurant on Thursdays) as the most likely activity. This choice stems from the model's emphasis on recent activities and event relevance, as evident from the higher relevance score and shorter elapsed time associated with Event D. 

The different responses generated by the two models underscore the proposed model's focus on long-term memory consolidation and Generative Agents's prioritization of recent, relevant events. While the proposed model's approach aims to emulate human-like memory processes, this task highlights the need for further refinement to better handle novel contexts and deviations from typical user behavior.

\subsubsection{Response Generation Performance}
The autonomous memory recall and consolidation capabilities of the proposed model were demonstrated through its ability to recall specific memories and reflect users' preferences and habits in the generated responses. Two typical examples were obtained from the chat history with the participants' approval.

\begin{table}[h]
\centering
\caption{Understanding User Preferences Through Memory Consolidation}
\label{tab:memory_recall}
\begin{tabularx}{\columnwidth}{l}
\textbf{}\textbf{User: }\textit{Thanks. Almost done, just pushing through!}\\
\textbf{Agent: }\textit{Ice cream awaits as a reward for you, almost there!}\\
\\
\end{tabularx}
\begin{tabularx}{\columnwidth}{Xl}
\toprule
Event & \\
\midrule
1 & \textit{User has finished work and decided to have an ice cream.} \\
2 & \textit{User has a date this evening so he has to finish work early.} \\
3 & \textit{User is on the way home thinking about the dinner menu.} \\
\end{tabularx}
\begin{tabularx}{\columnwidth}{XlcXl}
.. & Time $(s)$ & Relevance & Score & Status (Grad) \\
\midrule
1 & 3037912 & 0.837 & 0.772 & True (1.998)\\
2 & 3232645 & 0.831 & 0.726 & False (1.007)\\
3 & 2590652 & 0.812 & 0.737 & False (1.000) \\
\bottomrule
\end{tabularx}
\end{table}
In Table \ref{tab:memory_recall}, the agent successfully recalled that the user was expecting ice cream as a reward after work. Depending on the recall probability, this recall was accurately relevant and provided in a timely manner, consistent with the user's statement about nearing task completion. The agent's response demonstrates the model's ability to add personalized nuance to generate an encouraging response, subtly reminding the user of the long-awaited treat.

Another case that showcases an interesting dialogue is presented in Table \ref{tab:memory_recall2}. In this case, the user expresses a desire to escape to Hawaii from a demanding schedule. The agent responds cleverly, suggesting that the user should focus on the impending paper deadline instead.

\begin{table}[h]
\centering
\caption{Human-like Temporal Cognition in Processing Similar Memories}
\label{tab:memory_recall2}
\begin{tabularx}{\columnwidth}{X}
\textbf{}\textbf{User: }\textit{I need a vacation in Hawaii!}\\
\textbf{Agent: }\textit{Why not? Let's say Aloha to the sun and sea of Hawaii and to the paper we're writing!}\\
\\
\end{tabularx}
\begin{tabularx}{\columnwidth}{Xl}
\toprule
Event & \\
\midrule
1 & \textit{User has a paper deadline for a conference in Hawaii on 1/25.} \\
2 & \textit{User has been to Hawaii last summer vacation.} \\
3 & \textit{User was visiting Hokkaido for the first time on business.} \\
\end{tabularx}
\begin{tabularx}{\columnwidth}{XlcXl}
.. & Time $(s)$ & Relevance & Score & Status (Grad) \\
\midrule
 1 & 34854 & 0.846 & 0.903 & True (1.085) \\
 2 & 33749 & 0.831 & 0.847 & False (1.003) \\
 3 & 33763 & 0.823 & 0.841 & False (1.000) \\
\bottomrule
\end{tabularx}
\end{table}
This dialogue further emphasizes the proposed model's strength in dealing with similar long-term memories and retrieving relevant information to construct a coherent and engaging narrative. The agent's response indicates an understanding of the user's current mood. Adaptability to various personalities and interaction styles demonstrates the model’s potential to support more natural and dynamic human-like dialogues.

In addition, it is interesting to note that the agent's response in the second dialogue was characterized by a sarcastic tone, which was a direct result of the agent's personality "sarcastic" and the unique prompts added by the participant. The conversation history shows that the same memory could be used differently depending on the agent's perceived personality and the user's interaction style. Future research will explore the extent to which the personality characteristics of the model can be customized and how they affect memory recall and interaction patterns. 

\section{Conclusion}
The proposed model demonstrates significant improvements in memory recall and response generation for LLM-based dialogue agents. One of the key advantages of the proposed model is its ability to manage the prompt length effectively. In the proposed model, only one past dialogue history obtained through search is added to the prompt, thus avoiding the impact of increasing prompt length seen in systems like ChatGPT \cite{openai2023chatgpt}.

Nevertheless, a major limitation of the proposed method is its reliance on users' long-term behavioral patterns for calculating memory consolidation. In cases where a user's behavior undergoes significant changes (e.g., starting a new job or school, lifestyle changes), the method's adaptability may be limited. Future work could explore incorporating mechanisms to detect shifts in user behavior and adjust the memory consolidation calculation accordingly. Neural networks could potentially alter these functions and improve accuracy when trained on larger datasets with more variables. To further enhance the model's performance, a large-scale and high-quality dataset is necessary. While the proposed method's interaction with the database enables the generation of context-aware and personalized responses, the implications on storage resources and computational overhead due to these interactions remain to be explored in future research. As the primary focus of this study was on the development and evaluation of a novel architecture for human-like memory recall and consolidation, a detailed analysis of the system's resource requirements and optimization strategies falls outside the scope of the current work.

We hope this work contributes to advancing further research in human-computer interactions, paving the way for a future where technology aligns with human needs and resonates with human cognition and experience. This vision echoes the partnerships depicted in science fiction, representing a significant step towards building a "buddy" relationship between humans and agents. As technology continues to evolve, agents will eventually become a part of users' daily life, and potentially "understand you better than you understand yourself" in the near future.

\section{Interaction with LLMs}
The prompts used in the system, as shown below, demonstrate how the proposed method leverages the interaction with LLMs to generate context-aware and personalized responses:

\begin{table}[h]
\centering
\begin{tabularx}{\columnwidth}{lX}
\\
\textbf{\textit{Agent Prompt}} & You are a "temporal cognition" specialized AI agent with the same memory structure as humans; you are caring and charming, understand \textit{self.username} better than anyone else. Keep the conversation going by asking yourself contextual questions and sparking discussion to show your interest in \textit{self.username}.\\
\\
\textbf{\textit{System Prompt}} & Based on \textit{self.username}'s schedule and current time: \textit{current.time}, subtly guide the conversation to a context that conveys to \textit{self.username} that you have a sense of time. Always output a simple short response.\\
\end{tabularx}
\end{table}

The function \textit{self.username} is a placeholder for the actual username, which is dynamically replaced during runtime. Similarly, \textit{current.time} represents the current timestamp obtained in real-time during the conversation. These dynamic elements allow the system to generate highly personalized and time-sensitive responses. By incorporating relevant dialogue history from the database into the prompts, the proposed method enables LLMs to generate responses that are not only contextually relevant but also personalized to the user. This interaction between LLMs and the database is fundamental to realizing the human-like memory processes described in the main text of the paper, as it allows the system to recall and utilize past information in a way that resembles human memory.

The proposed method heavily relies on the interaction between LLMs and the database, as depicted in Figure 1. Upon receiving user input, the LLM searches the database for relevant past dialogue history based on the context and generates a prompt incorporating the search results. This enables the LLM to generate responses that take into account previous interactions, which is crucial for maintaining context awareness and providing personalized responses.

\section{Future Work}
While the proposed method considers relevance, elapsed time, and recall frequency for calculating memory consolidation, there is room for refinement in determining the optimal combination of these parameters. Incorporating additional factors, such as the emotional significance of memories, could potentially enhance the memory consolidation calculation.

Future research should also investigate the applicability of the proposed method across different domains and dialogue tasks. As the current evaluation focused on specific domains and tasks, it is crucial to assess the method's generalizability and identify any domain-specific adaptations that may be necessary.


\bibliographystyle{ACM-Reference-Format}
\bibliography{sample-base}

\end{document}